\documentclass{amsart}

\begin{document}

\title{A type of simulation which some experimental evidence suggests we don't live in}

\author{Samuel Alexander}

\date{July 1st, 2018 (Published in \emph{The Reasoner} 12 (7), p.\ 56)}

\begin{abstract}
Do we live in a computer simulation?
I will present an argument that the results of a certain
experiment constitute empirical evidence that we do not live
in, at least, one type of simulation. The type of simulation
ruled out is very specific. Perhaps that is the price one must
pay to make any kind of Popperian progress.
\end{abstract}

\maketitle

Published in \emph{The Reasoner} 12(7), p.\ 56 (2018).

%2%
In electronics, a \emph{soft error} is a type of error caused by a particle
hitting a computer's memory banks.
Early computer chips were manufactured with materials that
emitted alpha particles due to radioactive decay. These alpha
particles could hit memory cells and change memory values.
The same phenomenon can happen if a cosmic ray hits the computer.

%3%
Suppose we live in a computer simulation with the following \emph{$x$-$\hat{x}$
property}: for each memory-bit $x$ in any computer in our world,
there is a memory-bit $\hat{x}$ in the simulating computer
such that $\hat{x}$ is used to store which value is stored in
$x$. Then any such $x$ is subject to two different types of soft errors:
\begin{itemize}
    \item (Internal) Soft errors caused by simulated particles hitting $x$ in our simulated universe.
    \item (External) Soft errors caused by real particles hitting $\hat{x}$ in the universe where
        the simulation takes place.
\end{itemize}
Further, assume the following \emph{uni-directional property}:
putting a simulated memory-bit inside a simulated vault does not protect it
from external soft errors. We mean ``vault'' literally: a non-metaphorical
barrier of hard matter in the simulated universe.

%4%
Putting a simulated memory-bit in a simulated vault might protect it from internal soft errors, because
a thick vault might physically block incoming particles. The uni-directional property says this
defense cannot prevent external soft errors. If we live in an external-soft-error-prone simulation with the
$x$-$\hat{x}$ property and the uni-directional property, no vault we build can perfectly protect
a memory-bank from all soft errors, because each memory-bit $x$ in that memory-bank remains susceptible
to external soft errors caused by real particles hitting $\hat{x}$.

%5%
A paper by O'Gorman et al [1]]
describes (p.\ 46) the
following experiment and its results. A total of 864 modules were first run on the
second floor of a two-story building for 4,671 hours, during which time, 24 soft errors
were detected. Then, the same 864 modules were run for 5,863 hours in a nearby vault
shielded by about 20m of rock, during which time, zero soft errors were detected.

%6%
The above results suggest that a vault of 20m of rock blocked all soft errors.
By the above remarks, this is experimental evidence that we do not live in an
external-soft-error-prone simulation with the $x$-$\hat{x}$ property and the uni-directional
property. If we do live in such a simulation, then it should not
be possible to protect a simulated memory-bank with a simulated vault.

%7%
Of course, this is not a mathematical proof, merely empirical evidence.
The evidence could be improved, or the thesis falsified, with further experiments.
What if we repeat the experiment and soft errors are detected in the vault?
Without additional technology, we are unable to tell which soft errors were
external and which were internal.
(We can only distinguish them vacuously: if zero soft errors occur, then
zero are external and zero are internal.)
If soft errors
persist in settings more and more hostile to internal-soft-errors,
that is evidence that either we're overlooking (and failing to control for)
some unknown source of internal soft errors, or else that external soft errors exist.
The latter would entail we live in an external-soft-error-prone simulation,
albeit not necessarily one with the
$x$-$\hat{x}$ and uni-directional properties (maybe $x$-$\hat{x}$ fails
for other memory-bits besides the ones tested; maybe soft errors in the simulating
computer affect non-memory components of the simulated computers, indirectly manifesting
as soft errors in simulated computers; and so on).

%8%
Perhaps this paper's most interesting conclusion is just that a non-contrived simulation hypothesis is
falsifiable in a concrete way.
One can easily imagine many types of simulations we could live in that are not external-soft-error-prone,
or that lack the $x$-$\hat{x}$ property, or that lack the uni-directional property.
I hope my argument will inspire falsifiable predictions of
other types of simulations.

\end{document}